# PERFORMANCE ANALYSIS OF AODV, DSDV AND DSR IN MANETS

Akshai Aggarwal[1,], Savita Gandhi[2], Nirbhay Chaubey[3]

[1] Gujarat Technological University, Ahmedabad, Gujarat, India
`vc@gtu.ac.in`

[2] Department of Computer Science, Gujarat University, Ahmedabad, Gujarat, India
`drsavitagandhi@gmail.com`

[3] Department of Computer Science, Institute of Science and Technology for Advanced Studies and Research, Vallabh Vidyanagar, Gujarat, India
`nirbhay@ieee.org`

**ABSTRACT**

*Mobile Ad hoc Networks (MANETs) are considered as a new paradigm of infrastructure-less mobile wireless communication systems. MANETs are being widely studied and it is the technology that is attracting a large variety of applications. Routing in MANETs is considered a challenging task due to the unpredictable changes in the network topology, resulting from the random and frequent movement of the nodes and due to the absence of any centralized control [1][2]. In this paper, we evaluate the performance of reactive routing protocols, Ad hoc On demand Distance Vector (AODV) and Dynamic Source Routing (DSR) and proactive routing protocol Destination Sequenced Distance Vector (DSDV). The major goal of this study is to analyze the performance of well known MANETs routing protocol in high mobility case under low, medium and high density scenario. Unlike military applications, most of the other applications of MANETs require moderate to high mobility. Hence it becomes important to study the impact of high mobility on the performance of these routing protocols. The performance is analyzed with respect to Average End-to-End Delay, Normalized Routing Load (NRL), Packet Delivery Fraction (PDF) and Throughput. Simulation results verify that AODV gives better performance as compared to DSR and DSDV.*

## 1. INTRODUCTION

Mobile ad hoc networks (MANETs) are a heterogeneous mix of different wireless and mobile devices, ranging from little hand-held devices to laptops that are dynamically and arbitrarily located in such a manner that the interconnections between nodes are capable of changing on a continual basis [1].

An ad hoc network is a group of wireless mobile computers (or nodes) in which nodes cooperate by forwarding packets for each other to allow a node to communicate beyond its direct wireless transmission range. Ad hoc networks require no centralized administration or fixed network infrastructure such as base stations or access points and can be quickly and inexpensively set up as needed. In Ad Hoc Networks the individual mobile hosts (nodes) act at the same time as both the router and the host.

In a MANET, nodes within each other's wireless transmission ranges can communicate directly. However when a node wants to send a message to another node, which is situated outside its communication range, it has to rely on some other nodes to relay its messages [3]. Thus, a multi-hop scenario occurs, where several intermediate hosts relay the packets sent by the source host before they reach the destination host. The network topology may change with time as the nodes move or adjust their transmission and reception parameters.

Routing protocols are used to find routes for transmission of packets. Routing is the most fundamental research issue in MANETs. The merit of a routing protocol can be analyzed through metrics-both qualitative and quantitative. Desirable qualitative properties of a routing protocol for MANETs are Distributed operation, Loop-freedom, Demand-based operation, Security, Sleep period operation and unidirectional link support. Some quantitative metrics that can be used to assess the performance of any routing protocol are End-to end delay, throughput, PDF, NRL and Route Acquisition Time etc.

Routing protocols for ad hoc networks must deal with limitations such as high error rates, scalability, security, quality of service, energy efficiency, multicast, aggregation and node cooperation etc.

This paper is structured as follows: In Section 2, we discuss some of the routing protocols used in MANETs. Section 3 discusses related work. Performance metrics for routing protocols, used in MANETs, are described in section 4. The Simulation set- up is given in section 5. The results are discussed in section 6. The last section presents the concluding remarks.

## 2. ROUTING PROTOCOLS FOR MANETs

Routing protocols for wireless ad hoc networks can be mainly classified into the two categories: Table-driven (or Proactive) and On-demand (or Reactive) [3].

### 2.1 Pro-active Routing (Table-driven)

Table driven ad hoc routing protocols maintain at all times routing information regarding the connectivity of every node to all other nodes that participate in the network. Also known as proactive, these protocols allow every node to have a clear and consistent view of the network topology by propagating periodic updates. Therefore, all nodes are able to make immediate decisions regarding the forwarding of a specific packet. The main disadvantages of such algorithms are –
  i. Requirement for maintenance of a large amount of data at every node.
  ii. Slow reaction on restructuring and failures.

**2.1.1 Destination Sequenced Distance Vector (DSDV)**

This algorithm uses routing table like Distance vector but each routing table entry is tagged with sequence number, generated by destination. To maintain consistency among routing tables in a dynamically varying topology, updates are transmitted periodically. Each mobile station advertizes its own routing table to its current neighbors [4].

Routing information is advertised by broadcasting or multicasting. Packets are transmitted periodically and incrementally as changes are detected. In a wireless medium broadcasts are limited by the physical characteristic of medium. If a node invalidates its entry to a destination due to loss of next hop node, it increments its sequence number and uses new sequence number in its next advertisement of the route. Data broadcast by each mobile computer will contain new sequence number and

   i.   Destination IP address
   ii.  Number of hops required to reach the destination
   iii. Sequence number of the information received regarding that destination

To reduce the information carried in each broadcast message, two methods exist
 i. Full dump: The dump carries all the available routing information
 ii. Incremental carry: The message carries only changed information since the last full dump.

It may happen that every time a mobile host receives a worse metric than the upcoming sequence number update. In that case, route to the destination may change at every new sequence number. Solution to this problem is to delay the advertisement if mobile host can determine that a route with a better metric is likely to show up soon. For this two routing tables are maintained, one for forwarding packets and the other for incremental routing information packets. DSDV guarantees a loop free path to each destination without requiring nodes to participate in any complex update coordination protocol. In this protocol, routing tables of each node can be visualized as forming N trees, one rooted at each destination [4].

DSDV is one of the early algorithms available and the main advantage of this protocol is that it is quite suitable for creating ad hoc networks with a small number of nodes. One of the disadvantages of this protocol is that it requires a regular update of its routing tables, which uses up battery power and some amount of bandwidth, even when the network is idle. Secondly, whenever the topology of the network changes, a new sequence number is necessary before the network re-converges. Thus, DSDV is not suitable for highly dynamic networks.

## 2.2 Reactive Routing (On-demand)

Reactive routing protocols, which appear to be more suitable for ad hoc networks, do not maintain up-to-date information about the network topology, as is done by the proactive ones, but they create routes on demand. Among reactive routing protocols, the Ad hoc On Demand Distance Vector Routing (AODV) and the Dynamic Source Routing (DSR) are the most established and popular. This type of protocols finds a route on demand by flooding the network with Route Request packets.

### 2.2.1 Ad hoc On Demand Distance Vector (AODV)

This protocol performs Route Discovery using control messages Route Request (RREQ) and Route Reply (RREP). In AODV, routes are set up by flooding the network with RREQ packets which, however, do not collect the list of the traversed hops. Rather, as a RREQ traverses the network, the traversed mobile nodes store information about the source, the destination, and the mobile node from which they received the RREQ. The later information is used to set up the reverse path back to the source. When the RREQ reaches a mobile node, that knows a route to the destination or the destination itself, the mobile node responds to the source with a packet (RREP) which is routed through the reverse path set up by the RREQ. This sets the forward route from the source to the destination. To avoid overburdening the mobiles with information about routes which are no longer (if ever) used, nodes discard this information after a timeout. When either destination or intermediate node moves, a Route Error (RERR) is sent to the affected source nodes. When source node receives the RERR, it can reinitiate route discovery if the route is still needed. Neighborhood information is obtained by periodically broadcasting

Hello packets [5]. For the maintenance of the routes, two methods can be used: a) ACK messages in MAC level or b) HELLO messages in network layer.

The main advantage of this protocol is that routes are established on demand and destination sequence numbers are used to find the latest route to the destination. The connection setup delay is lower. One of the disadvantages of this protocol is that intermediate nodes can lead to inconsistent routes if the source sequence number is very old and the intermediate nodes have a higher but not the latest destination sequence number, thereby having stale entries. Also multiple RREP packets in response to a single RREQ packet can lead to heavy control overhead. Another disadvantage of AODV is that the periodic beaconing leads to unnecessary bandwidth consumption.

### 2.2.2 Dynamic Source Routing (DSR)

In DSR, when a mobile (source) needs a route to another mobile (destination), it initiates a route discovery process which is based on flooding. The source originates a RREQ packet that is flooded over the network. The RREQ packet contains a list of hops which is collected by the route request packet as it is propagated through the network. Once the RREQ reaches either the destination or a node that knows a route to the destination, it responds with a RREP along the reverse of the route collected by the RREQ [6]. This means that the source may receive several RREP messages corresponding, in general, to different routes to the destination. DSR selects one of these routes (for example the shortest), and it maintains the other routes in a cache. The routes in the cache can be used as substitutes to speed up the route discovery if the selected route gets disconnected. To avoid that RREQ packets travel forever in the network, nodes, that have already processed a RREQ, discard any further RREQ bearing the same identifier.

The main difference between DSR and AODV is in the way they keep the information about the routes: in DSR it is stored in the source while in AODV it is stored in the intermediate nodes. However, the route discovery phase of both is based on flooding. This means that all nodes in the network must participate in every discovery process, regardless of their potential in actually contributing to set up the route or not, thus increasing the network load.

## 3. RELATED WORK

To evaluate the performance of the routing protocols Chenna R. et al. [8], Talooki and Ziarati [9] and Lakshmikant et al. [10] presented a detailed simulation of DSDV, AODV, DSR and TORA with 50 wireless nodes forming ad hoc networks and the paper concluded that DSDV and TORA show good performance in a network with low mobility whereas AODV and DSR maintain comparatively better performance in all mobility situations. Mahdipur E, et. Al [11] evaluated the performance of DSDV and AODV routing protocols in MANETs under CBR traffic with NS-2 [7].

Performance comparison of AODV and DSR routing protocols in a constrained situation is done in [12]. The authors claim that the AODV outperforms DSR in normal situation but in the constrained situation DSR out performs AODV, where the degradation is as severe as 30% in AODV whereas DSR degrades marginally as 10%. Though both AODV and DSR use on demand route discovery, they have different routing mechanics. Perkins et all [13] observe that, for application oriented metrics such as delay and throughput, DSR outperforms AODV when the numbers of nodes are smaller. AODV outperforms DSR when the number of nodes is very large. The authors show that DSR consistently generates less routing load than AODV.

## 4. PERFORMANCE METRICS

In this paper, we consider following four performance metrics to compare the three routing protocol.

1. Average End-to-End Delay: It is defined as the average time taken by the data packets to propagate from source to destination across a MANET. This includes all possible delays caused by buffering during routing discovery latency, queuing at the interface queue, and retransmission delays at the MAC, propagation and transfer times.

2. Normalized Routing Load (NRL): The number of routing packets transmitted per data packet delivered at the destination.

3. Packet Delivery Fraction (PDF): This is the ratio of the number of data packets successfully delivered to the destinations to those generated by sources. Packet Delivery Fraction = received packets/sent packets * 100

4. Throughput: It is the rate of successfully transmitted data packets in a unit time in the network during the simulation.

## 5. SIMULATION SETUP

The simulations were performed using Network Simulator 2 (NS-2.33) [7]. The traffic sources are Constant Bit Rate (CBR). The source destination pairs are spread randomly over the network. The mobility model uses 'random waypoint model' in a rectangular field of 1000m x 1000m with 25 nodes to 200 nodes. Different network scenario for different number of nodes for 5 connections and 10 connections are generated. In Table 1, we have summarized the model parameters that have been used for our experiments.

| Parameter | Parameter Value |
| --- | --- |
| Simulator | NS-2.33 |
| Simulation Area | 1000m X 1000m |
| MAC Protocol | IEEE 802.11 |
| Mobile Nodes | 25,50,75,100,125,150,175,200 |
| Antenna Type | Omni antenna |
| Propagation Model | Two Ray Ground |
| Number of Connections | 5,10 |
| Packet Size | 512 byte |
| Routing Protocols | AODV, DSDV & DSR |
| Traffic Sources | CBR (UDP) |
| Simulation Time | 100 Sec. |
| Mobility Model | Random waypoint |
| Pause Time | 0 |

Table 1: Simulation Parameters

## 6. RESULTS AND DISCUSSION

In this Section, we compare the capabilities of the three routing protocol studied in this paper. To evaluate more reliable performance of AODV, DSDV and DSR routing protocols in same simulation environment (25 to 200 mobile nodes). Simulations results are collected from a total of 60 scenarios of the three protocols. Performance metrics are calculated from trace file, with the help of AWK program. The simulation results are shown in the following section in the form of line graphs. Graphs show comparison between the three protocols by varying different numbers of sources.

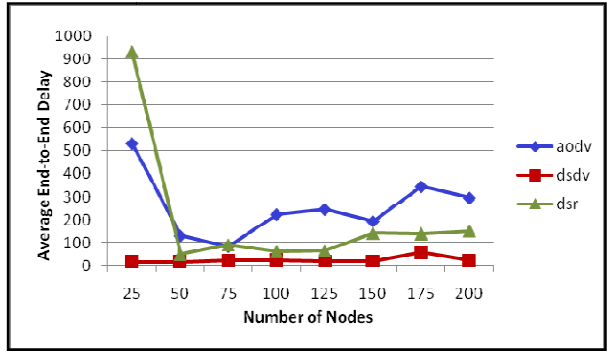

Figure 1. Average End-to-End Delay vs. Number of Nodes (with 5 Connections)

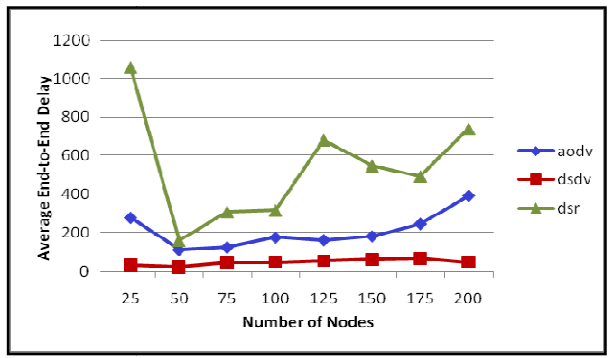

Figure 2. Average End-to-End Delay vs. Number of Nodes ( with 10 Connections)

The delay is affected by high rate of CBR packets as well. The buffers become full much quicker, so the packets have to stay in the buffers for a much longer period of time before they are sent. In Figure 1 DSR decreases and varies with the number of nodes in the networks, however, the performance of AODV is degrading due to increase in the number of nodes.

In Figure 2, we noticed that the performance of DSR is degrading due to increase in the number of nodes in the networks. The performance of the AODV is slightly better. Average delay is less for DSDV routing protocol and remains constant as the number of nodes increases.

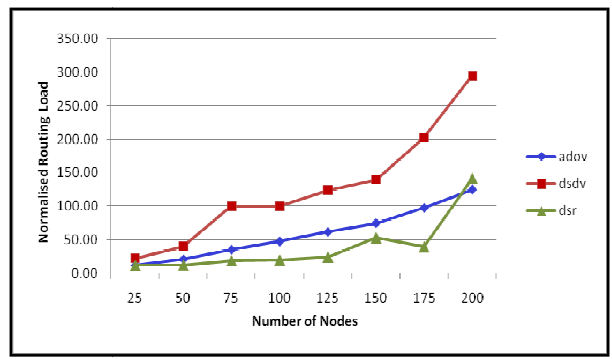

Figure 3. Normalized Routing Load Vs. Number of Nodes( with 5 Connections)

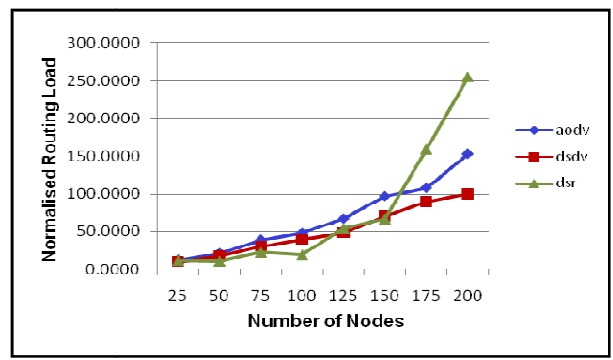

Figure 4. Normalized Routing Load vs. Number of Nodes ( with 10 Connections)

Normalized routing load (NRL) of AODV, DSDV and DSR protocols in different sources are presented in Figure 3 and Figure 4. In Figure 3 (5 connection/source), AODV and DSR demonstrate lower routing load. Proactive routing protocol DSDV showed higher routing load than the reactive routing protocols AODV and DSR. In Figure 4 (10 connection/source), as network load is increased, Normalized Routing Load of AODV and DSR is much higher than the DSDV. In this simulation, due to high congestion in the ad-hoc network, AODV requires more routing packets to maintain transmission of data packets. We have used the same simulation environment path, mobility and traffic patterns for these three protocols and AODV has consistent and worse NRL as the number of nodes is increased.

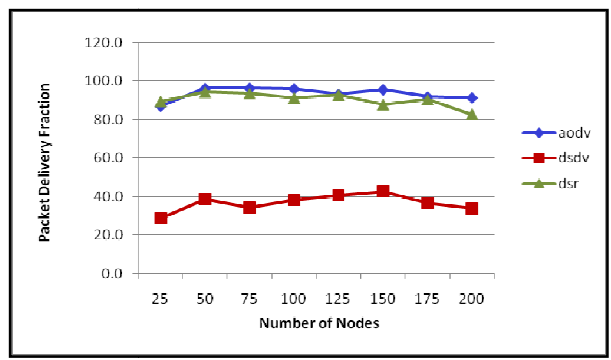

Figure 5. Packet Delivery Fraction Vs. Number of Nodes (with 5 Connections)

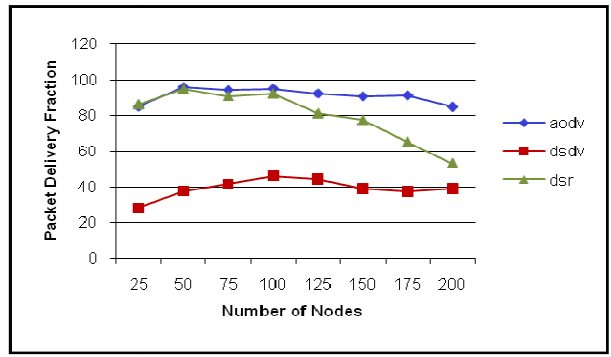

Figure 6. Packet Delivery Fraction Vs. Number of Nodes (with 10 Connections)

In Figure 5, we have noticed a slight advantage to AODV when the number of nodes is increased in mobile networks. Overall, the data packet delivery ratio of AODV and DSR is higher in a scenario with high mobility than that of DSDV.

Figure 6 shows that the AODV manages to deliver a greater fraction of data packets in scenarios with high mobility in large mobile networks. We observe that DSR routing protocol performs well when the number of nodes is less, however its performance declines drastically with increased number of nodes in the network. The performance of DSDV is better when the number of nodes is increasing in the network.

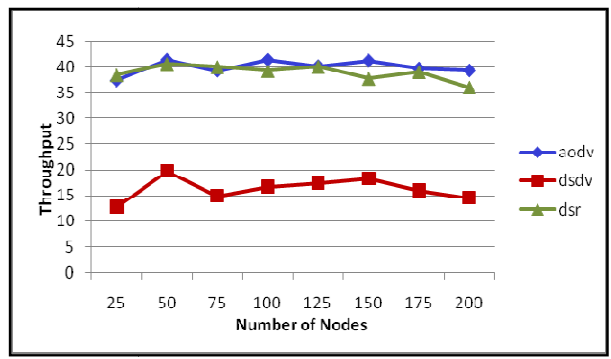

Figure 7. Throughput Vs. Number of Nodes (with 5 Connections)

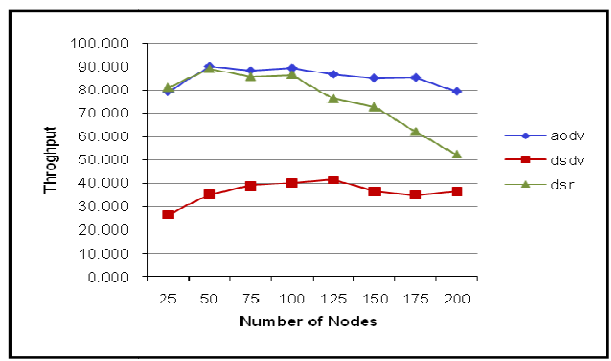

Figure 8.  Throughput Vs. Number of Nodes (with 10 Connections)

From the above Figure 7 and Figure 8 it is clear that AODV gives better throughput and outperforms even the DSR.

|      | Average End-to-End Delay | NRL | PDF | Throughput |
|------|--------------------------|-----|-----|------------|
| AODV | Performance Degrade with number of nodes increase in the networks | Consistent and worse NRL when increasing number of nodes. | Best | Best |
| DSDV | Least and remains constant as the number of nodes increase in the networks | Higher routing load than the AODV and DSR. | Least | Least |
| DSR  | Degrade when number of nodes increase in the networks. | Much higher than the AODV when network load is increased. | Performs well when the number of nodes is less but it declines drastically when the numbers of nodes are increased. | Better than DSDV |

Table 2: Result Analysis

## 7.  CONCLUSION AND FUTURE WORK

Our simulation work illustrates the performance of three routing protocols AODV, DSR and DSDV. The paper presents a study of the performance of routing protocols, used in MANETs, in high mobility case under low, medium and high density scenario. We vary the number of nodes from 25 (low density) to 200 (high density) in a fixed topography of 1000 x 1000 meters. Moreover, since Random Waypoint Mobility Model has been used in this study to generate node mobility, we take an average of 10 randomly generated scenarios so to make a detailed performance analysis. We find that the performance varies widely across different network sizes and results from one scenario cannot be applied to those from the other scenario. AODV performance is the best considering its ability to maintain connection by periodic exchange of information. As far as Throughput is concerned, AODV and DSR perform better than the DSDV even when the network has a large number of nodes. Overall, our simulation work shows that AODV performs better in a network with a larger number of nodes whereas DSR performs better when the number of nodes is small. Average End-to-End Delay is the least for DSDV and does not change if the no of nodes are increased. Thus, we find that AODV is a viable choice for MANETs but NRL for AODV increases at a higher rate compared to that in DSDV & DSR with increase in number of nodes in networks. In this paper, we have done complete analysis of the three MANET's routing protocols. We feel that the conclusion that we have reached about the performance are one of the most definitive comparison obtained by any researcher. Our future plan is to evaluate security issues in AODV.

## Authors

**Akshai Aggarwal** (MIEEE'1966, SMIEEE'1992) is working as Vice Chancellor, Gujarat Technological University, Ahmedabad, India. Before joining as the Vice-Chancellor, he was working as the Director of School of Computer Science, University of Windsor, Canada. He worked as Professor and Head of Department of Computer Science at Gujarat University for about 10 years. Before that he was Professor and Head, Department of EE at M.S.University of Baroda. He was Chairman of IEEE India Council for two years. He initiated IEEE activities in Gujarat by starting the first IEEE Student Branch at M.S.University of Baroda. Later he initiated the establishment of the Student Branch at Gujarat University. He was also the founder Chairman of IEEE Gujarat Section, the IEEE Computer Society Chapter and the IEEE Joint Chapter of Industry Applications, Industrial Electronics and Power Electronics. The Section conducted two International Conferences and one national Seminar during his Chairmanship. He graduated with a B.Sc.(EE) from Punjab Engg College and studied at MS University of Baroda for his Master's and Doctoral work. 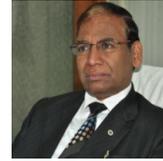

**Savita Gandhi** (MIEEE' 2003 SMIEEE' 2005) is Professor & Head at the Department of Computer Science, Gujarat University and Joint Director, K.S. School of Business Management, Gujarat University. She is with Gujarat University for about 20 years. Before that she has worked with M.S. University of Baroda, Department of Mathematics for about 10 years. She has been actively associated with IEEE activities at Gujarat Section. She is M.Sc. (Mathematics), Ph.D (Mathematics) and A.A.S.I.(Associate Member of Actuarial Society of India by the virtue of having completed the "A" group examinations comprising six subjects conducted by Institute of Actuaries , London). She was awarded Gold Medal for standing first class first securing 93% marks in M.Sc. and several prizes at M.Sc. as well as B.Sc. Examinations for obtaining highest marks. 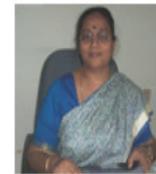

**Nirbhay Chaubey** (SIEEE' 2002 MIEEE' 2004) is working as an Assistant Professor of Computer Science at Institute of Science and Technology for Advanced Studies and Research, Vallabh Vidyanagar, Gujarat, India. Currently, he is pursuing Ph.D in Computer Science at Department of Computer Science, Gujarat University, Ahmedabad, India. He has been involved in IEEE activities since 1994. His position held for IEEE Gujarat Section include Executive Secretary (1998-2005), Treasurer (2005-2006), Secretary and Treasurer (2007) and Treasurer for year 2008 onwards. He graduated from Ranchi Unviersity, Ranchi, and Master in Computer Applications from Madurai Kamraj University, Madurai, India of Baroda for his Master's and Doctoral work. 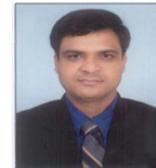